# Increased risk of type I errors for detecting heterogeneity of treatment effects in cluster-randomized trials using mixed-effect models

Noorie Hyun[a], Abisola E Idu[a], Andrea J. Cook[a] and Jennifer F. Bobb[a]

[a]Division of Biostatistics, Kaiser Permanente Washington Health Research Institute, Seattle, Washington

Correspondence: Noorie Hyun, Email: noorie.hyun@kp.org

**Abstract**

Evaluating heterogeneity of treatment effects (HTE) across subgroups is common in both randomized trials and observational studies. Although several statistical challenges of HTE analyses including low statistical power and multiple comparisons are widely acknowledged, issues arising for clustered data, including cluster randomized trials (CRTs), have received less attention. Notably, the potential for model misspecification is increased given the complex clustering structure (e.g., due to correlation among individuals within a subgroup and cluster), which could impact inference and type 1 errors. To illicit this issue, we conducted a simulation study to evaluate the performance of common analytic approaches for testing the presence of HTE for continuous, binary, and count outcomes: generalized linear mixed models (GLMM) and generalized estimating equations (GEE) including interaction terms between treatment group and subgroup. We found that standard GLMM analyses that assume a common correlation of participants within clusters can lead to severely elevated type 1 error rates of up to 47.2% compared to the 5% nominal level if the within-cluster correlation varies across subgroups. A flexible GLMM, which allows subgroup-specific within-cluster correlations, achieved the nominal type 1 error rate, as did GEE (though rates were slightly elevated even with as many as 50 clusters). Applying the methods to a real-world CRT using the count outcome utilization of healthcare, we found a large impact of the model specification on inference: the standard GLMM yielded highly significant interaction by sex (P=0.01), whereas the interaction was non-statistically significant under the flexible GLMM and GEE (P=0.64 and 0.93, respectively). We recommend that HTE analyses using GLMM account for within-subgroup correlation to avoid anti-conservative inference.

1. **Introduction**

Dealing with clustered data is extremely common in public health research. It can arise by design including cluster randomized trials (CRTs) [1,2] or due to the hierarchical structure of the data (e.g. multiple outcomes per participant, patients clustered within providers, or providers clustered within clinics). Addressing correlation of clustered outcome data in the analysis is critical to correctly estimate standard errors and avoid inflated type I error[1].

Model-based methods for analyzing clustered data can be broadly classified into conditional and marginal models according to the interpretation of the model parameters[2, 3]. Conditional models adjust for cluster-specific effects in estimating the intervention effect, and the model parameters are usually estimated by mixed-effects regression, such as generalized linear mixed models (GLMM). In contrast, marginal models, such as generalized estimating equations (GEE), are often employed when focusing on population-level effects and interpreting the intervention effect coefficient as a population-averaged effect.

Beyond estimating overall treatment or exposure effects within the study population, many studies also examine differences in effects across subgroups. These analyses are often referred to as *effect modification* or *heterogeneity of treatment effect (HTE)* analyses. For randomized controlled trials (RCTs), subgroup analyses may be conducted to satisfy regulatory requirements to evaluate whether efficacy is consistent across subgroups or to identify whether safety problems are constrained to specific subgroups[4, 5]. Alternatively, subgroup analyses may be conducted for exploratory purposes to identify subgroups most likely to benefit from an intervention[6]. Regression models are often used for HTE analysis that include the subgroup factor, the intervention group, and their interaction term(s), as well as any additional pre-specified covariates[7].

Most RCT literature on statistical approaches for HTE analyses addresses analytical issues observed in individually randomized trials without clustering. These issues include lack of power, multiple testing and the potential for false positive findings, and the limited interpretation of the marginal subgroup analyses from the standpoint of patients belonging to a combination of subgroup factors (Hemmings, 2014). In contrast, analytical issues regarding subgroup analyses for clustered data have received comparatively little attention. The correlation structure in which subgroups are nested within clusters (see **Figure**) could lead to statistical challenges in detecting effect modification across subgroups using model-based methods. We therefore sought to examine the impact on inference of the choice of modeling approach (GEE and GLMM with different correlation structures) on statistical inference.

This work was motivated by the Primary Care Opioid Use Disorder treatment (PROUD) trial, a pragmatic, two-arm cluster-randomized trial in 12 clinics that tested whether implementing the Massachusetts Model of nurse care management for opioid use disorder (OUD)—a clinic-level intervention—increased OUD treatment in diverse primary care settings in the United States[8]. In pre-planned HTE analyses of the main effectiveness outcome, which used a simple GLMM with only a random intercept for cluster, the study estimated similar, almost identical, within-subgroup intervention effects with large 95% CI, but extremely small yet statistically significant interaction effects between intervention and subgroups. This motivated the work for this paper to explore how this paradoxical conclusion could occur, by examining whether the misspecification of the correlation structure may have led to the results, and to propose a more flexible approach for future analyses.

In this paper, we examine the possibility that model misspecification of routine analytic methods for clustered data may lead to type I error inflation when testing for the presence of effect modification across subgroups. We consider the scenario in which the within-cluster correlation differs across the subgroups being examined, and we propose a flexible GLMM allowing cluster-nested subgroups' correlation in Section 2. To examine the impact of model (mis)specification, we conduct a simulation study in which we compare results of both GLMM and GEE models with different correlation structures and across different outcome types, including continuous, binary, and count outcomes, in Section 3. Application to the motivating PROUD study is presented in Section 4. Finally, we provide recommendations and additional discussion in Section 5.

## 2. Statistical methods for effect modification across subgroups in cluster-randomized trials

We consider the setting of a parallel-group cluster RCT with two arms, $I$ clusters and cluster size $n_i$ for cluster $i$. Observed are outcome measurements $Y_{ij}$ from subject $j$ within cluster $i$, binary indicator $trt_i$ for the assigned intervention or control to cluster $i$, and a vector of (p-1) dummy variables for a subgroup factor $\boldsymbol{gr}_{ij}$. For simplicity, we assume a simple unadjusted model including only the subgroup of interest, but additional covariates could be straightforwardly incorporated as desired.

### 2.1. GEE and GLMM

We first present two traditional models applied to cluster RCTs. For GEE, the mean function is given by

$$g\{E(Y_{ij})\} = \beta_0 + trt_i * \beta_{trt} + \boldsymbol{gr}_{ij}^T \boldsymbol{\beta}_{gr} + (trt_i * \boldsymbol{gr}_{ij})^T \boldsymbol{\beta}_{mod}. \quad (1)$$

We consider canonical links for $g(\cdot)$, that is, the identity, log and logit functions for continuous, count and binary outcomes, respectively. The regression coefficient $\beta_{trt}$ is the intervention effect within the reference subgroup, and $\boldsymbol{\beta}_{mod}$ is the vector parametrizing the difference in intervention effects comparing the other subgroup categories to the reference subgroup (on the difference, log relative risk, or log odds ratio scale, depending on the link function). Any non-zero components of $\boldsymbol{\beta}_{mod}$ correspond to subgroups showing heterogenous intervention effects.

For an asymptotic variance estimator of GEE, we use the robust sandwich variance based on model-based information matrix and design-based empirical variance[9]. We used an independent working correlation in our simulation and application that is commonly used in practice to estimate an average intervention effect at the person-level[10]. Note that using other working correlation structures, such as exchangeable, may improve relative efficiency, but the advantage of GEE is the robustness to misspecified working correlation structures. By including GEE in our simulation, we sought to evaluate if using a simple correlation structure in practice achieved valid inferences in the HTE setting or if GEE would have similar issues as observed for GLMM.

For GLMM, the conditional mean function is similar to the GEE model above, but includes a cluster-specific random effect $U_i$:

$$g\{E(Y_{ij} \mid U_i)\} = \beta_0 + trt_i * \beta_{trt} + gr_{ij}^T \boldsymbol{\beta}_{gr} + (trt_i * gr_{ij})^T \boldsymbol{\beta}_{mod} + U_i, \quad (2)$$

where $U_i \sim N(0, \sigma^2)$ are i.i.d. with finite variance. To estimate the model parameters, marginal log-likelihoods are calculated by integrating out the random effects; for non-identity link functions no closed form exists such that approximate methods are employed[11]. For computational efficiency in our simulation study, we used the Laplace approximation to calculate the marginal log-likelihoods (default setting in the 'lmer4' R-package used to fit mixed models for this study[12]).

**2.2. Flexible GLMM Model**

We propose a more flexible GLMM that includes cluster-nested subgroup-specific random intercepts. For this model, the conditional mean function is given by:

$$g\{E(Y_{ij} \mid U_{i,s_{ij}})\} = \beta_0 + trt_i \boldsymbol{\beta}_{trt} + gr_{ij}^T \boldsymbol{\beta}_{gr} + (trt_i * gr_{ij})^T \boldsymbol{\beta}_{mod} + U_{i,s_{ij}}, (3)$$

where subgroup subscript $s_{ij}$ corresponds to the subgroup level of subject $j$ in cluster $i$ and $s_{ij} = 1, \ldots, p$, where $p$ is the number of subgroup levels. We assume that $\boldsymbol{U}_i \coloneqq (U_{i,1}, U_{i,2}, \ldots, U_{i,p})$ follows a multivariate normal distribution with zero mean and $\Sigma$ is a $p \times p$ covariance matrix, which is positive-definite. We further assume that the $\boldsymbol{U}_i$ are independent across clusters. It is also assumed that given $\boldsymbol{U}_{i,s_{ij}}$, $Y_{ij}$ is independent of $Y_{i'j'}$ for any $j$ and $j'$ given different clusters $i$ and $i'$. We consider canonical link functions as described in Section 2.1. We use the 'lmer4' R-package version 1.1-31 to fit the proposed model.

2.2.1 Singular Model Fitting Issues and Two-Step Procedure

Given the more complex correlation structure, the flexible GLMM model may be more likely to encounter a singular fit than the standard GLMM introduced in Section 2.1. One approach when a singular fit occurs is to simply proceed with the fitted model estimates. An alternate approach is to apply a two-step procedure. Under this two-step procedure, after encountering a singular fit (e.g., due to one of the subgroup-specific random effects having variance estimated as 0), one proceeds by fitting the standard GLMM from Section 2.1 that includes a common random intercept variance for all subgroups. We evaluate both approaches in the simulation study in the scenarios in which we observe singular fits occurring, including small random effects variances and/or small numbers of clusters.

Further, to avoid computational difficulties, in the simulation study in which the subgroup is only two-levels (e.g. $p=2$), we reparametrize the subgroup-specific random intercepts in flexible GLMM by using a random intercept and random slope parameterization for the subgroup dummy variables (see sample code in the **Appendix**). The reparametrized model reduces the occurrence of singularity issues defined by random effect estimates equal to 0. Although mathematically equivalent, we note that the two models that include subgroup specific random intercepts and random intercept and slope, respectively may not yield identical estimates due to the algorithm convergence. Although reparametrizing the flexible GLMM results in less frequent singular fit issues, given the more complex correlation structure the flexible GLMM is still more likely to yield a singular fit relative to the traditional GLMM model.

## 2.3 Small Cluster Size

For settings with a small number of clusters (< 50), various methods for correcting inferences have been studied in the setting of no subgroup-specific correlations within cluster for GLMM and GEE[13-16]. However, small cluster correction methods for flexible GLMM or for the setting of existing subgroup-specific correlations within cluster have not been systematically studied. Thus, we applied four existing correction methods to GLMM and flexible GLMM: (a) Satterthwaite degrees of freedom for the t-test for continuous outcomes, (b) subtracting the number of cluster-level parameters for fixed effects from the number of clusters [17] for the t-test statistics for binary and count outcomes, (c) "between-within" denominator degrees of freedom of freedom approximation for the t-test statistics for binary and count outcomes[18], and (d) parametric bootstrapping 95% confidence interval based on 100 simulated datasets from the fitted model. The 'parameters' R-package was used to implement corrections (c) and (d). Given the computational burden of bootstrapping within a simulation study, we first applied (a), (b) and (c) to the flexible GLMM and GLMM. If at least one of these methods yielded a type I error rate close to the nominal level (4%~5%), we presented the results from the best performer from among these three methods. However, if none of the methods maintained the nominal level, we additionally applied the bootstrapping method, and then chose the best performer (from among all four methods). Lastly, the Fay and Graubard correction was used in the GEE model[16, 19].

## 3. Simulation Study

We compared the proposed flexible GLMM (3), with and without the two-step procedure when encountering singular fits, with GEE (1) and GLMM (2). We primarily evaluated the methods in terms of the type I error rate for testing for effect modification between the intervention indicator and a subgroup factor (focusing on a binary subgroup variable for simplicity). We set the nominal significance level to be 5%. In addition to the type I error rate, we also compared bias and empirical standard deviations. For these latter comparisons, we focus on outcomes that are continuous or count data because the three models' regression coefficients can be comparable because identity and log link functions are collapsible. We considered four scenarios, Scenarios 1-4 as described below.

## 3.1 Simulation Set-up

We consider scenarios of cluster RCTs with N clusters, 1:1 random assignment of clusters to two treatment groups (e.g., intervention or control), and a balanced 2-level subgroup factor within a cluster. For data generation, we define a binary subgroup indicator $gr_{ij}$, (0 vs. 1), which is independently generated from a Bernoulli distribution with 0.5 probability. We independently generate $U_i \sim N(0, \sigma^2)$ if model (2) is the true model or $(U_{i,0}, U_{i,1})$ if model (3) is the true model from a bivariate normal distribution with zero mean and $\Sigma$ covariance. When $Y_{ij}$ is continuous, count and binary, we assume $Y_{ij}$ follows a normal distribution with an identity link, Poisson distribution with log link, and Bernoulli distribution with logit link, respectively.

Universal parameters used for all four scenarios are first described. Based on the estimated coefficients in our motivating PROUD dataset, we set the regression parameters $(\beta_0, \beta_{trt}, \beta_{gr}, \beta_{int})$ to be $(0, 0.5, 0.3, 0)$ for the continuous outcome setting, $(-1, -0.07, -0.5, 0)$ for count outcomes, and $(-1.66, -0.32, -0.08, 0)$ for binary outcomes. The residual error variance for continuous outcome is 0.64.

We considered four different data generating scenarios with parameter values shown in Table 1. These included three scenarios in which the flexible GLMM was the true model: an "ideal" setting in which there are a large number of clusters and moderately sized random effects (Scenario 1), as well as more challenging settings with either small random intercepts (Scenario 2) or a small number of clusters (Scenario 3). We also considered a scenario where the standard GLMM is the true data generating model (Scenario 4). Under each scenario, we repeated the simulation 1,000 times. For each generated dataset, we fit each of the comparator approaches described in Section 2. For testing for effect modification, we used a Wald test, with small sample correction methods under settings with 12 clusters as described above in Section 2. The selected correction methods for the scenarios with a small number of clusters (including Scenario 3 and 4) are summarized in Table 2.

Bias is calculated by the difference between the average of the point estimates and the true value; the empirical standard deviation is the standard deviation of the point estimates. Type I error rates are calculated as the proportion of p-values for effect modification that are less than the nominal 0.05 level. For the parametric bootstrapping method, the percentage of 95% confidence intervals excluding 0, the null effect modification, was calculated.

## 3.2 Simulation Results

The summary statistics of Scenarios 1-4 with 1,000 simulation replicates are presented in Tables 3-6, respectively. Note that no singular fit issue occurred in Scenario 1, so the flexible GLMM with the two-step procedure was not relevant. For continuous outcomes with an identity link in Scenario 1, the three models produce equivalent point estimates, resulting in similar bias magnitude and empirical standard deviations across the models (Table 3). However, the GLMM leads to an inflated type I error rate of up to 21.8% (>5%) due to the underestimation of the asymptotic standard error, resulting from covariance misspecification. In contrast, the flexible GLMM produces the nominal type I error rate, while the GEE model shows an acceptable type I error rate, particularly, when the number of clusters is 100. The empirical standard deviations of the regression coefficients estimates are similar in the flexible GLMM and GLMM. However, the GEE model shows slightly less efficiency. A similar pattern is observed for count and binary outcomes, with GLMM having an inflated type I error rate of up to 47.2% .

In Scenario 2 with small random effects (Table 4), the flexible GLMM is more likely to encounter singular fit issues than the GLMM (in up to 40% of simulation repetitions with count outcomes). The singularity issue is predominantly observed in the random slope estimates rather than the random intercept estimates. Increasing the number of clusters to 100 slightly reduces the rates of singularity across all three types of outcomes. The GLMM for binary outcome experiences a singular fit in a small number of simulations (75 and 28 when the number of clusters are 50 and 100, respectively) but not for continuous and count outcomes. Despite the singular fit, the point estimates of the three models are similar. However, when the number of clusters is 100, the flexible GLMM and GEE conservatively control the type I error rate for the null effect modification compared to GLMM across all three outcomes. Most of the type I error rates of the two-step procedure meet the nominal level. When the random effects are small, the degree of type I error inflation in fitting the GLMM is reduced but still present.

Scenario 3 reflects the data from the PROUD study (Table 5), specifically with 12 clusters of varying sizes. The small cluster correction methods described in Section 2 are applied. Compared to Scenario 2 across the three outcomes, the flexible GLMMs exhibit a decrease in the number of singular fits, and the GLMMs encounter no singular fit. Additionally, compared to GEE and GLMM, the flexible GLMM maintains the type I error rates close to the nominal level for the three outcome types and for both fixed and varying cluster sizes. In contrast, GEE has slightly inflated type I error (0.067-0.123) and the GLMM had very inflated type I error rates across all scenarios (0.138-0.449).

In Scenario 4 (Table 6), when the number of clusters is 50, with a fixed cluster size of 100, given that the GLMM is the true model in the data generation, GLMM effectively maintains the nominal type I error rate across the three outcomes. The flexible GLMM and GEE also control the type I error rate at the nominal level for continuous outcomes. However, for count and binary outcomes, the flexible GLMMs slightly deflate the type I error rate (0.036-0.039), whereas the GEEs slightly inflate the type I error rates (0.063-0.069). In the PROUD-like setting, the GLMMs yield the nominal type I error rate for continuous and count outcomes but yield a slightly lower type I error rate for binary outcomes. The flexible GLMMs slightly deflate the type I error rate, whereas the GEE slightly inflates the type I error rates across the outcome types.

## 4. Data Application

Here we re-analyze an HTE analysis from the PROUD study which evaluated whether implementation of office-based addiction treatment (intervention)—relative to usual care—reduced acute care utilization among primary care patients with opioid use disorder (OUD). The PROUD trial was funded by the National Institute for Drug Abuse (NIDA) Clinical Trials Network (CTN) and was conducted in 12 clinics from six diverse health systems across the US from 2015 to 2020. Within each health system, one of two clinics was assigned to the intervention and the other to usual care. The patient-level effectiveness outcome is the number of days of utilization of acute care services, including emergency department visits, urgent care visits, and inpatient hospitalizations. Details of study design and primary outcome results have been previously published[8].

For this application, we focus on examining the potential for modification of the intervention effect by sex as documented in patients' electronic health records. We employed flexible GLMM, GLMM, and GEE, using a Poisson distribution with log link to analyze the count outcomes. The GLMM accounted for the random intercept due to the 12 clinics, while the flexible GLMM accounted for the random intercept due to the clinics and the random slope due to sex (female vs. male). In all models, we adjusted for the baseline value of the outcome (reflecting the study's primary analysis approach). We applied the same correction techniques for small numbers of clusters (12 clinics) as described in Scenario 2, using a t-test with between-within correction for flexible GLMM, with 10 degrees of freedom (=12-2) for GLMM and GEE. Particularly, Fay and Graubard's correction for the GEE model couldn't be applied because of the singular information matrix. To align with the simulation study, we note that this reanalysis does not exactly replicate the primary outcome GLMM analysis, which used a likelihood ratio small-cluster test[14]. The 95% confidence intervals of the intervention effect within men and women were also small-cluster corrected by using the analogous methods used for p-value correction.

The two clinics assigned to the intervention and usual care groups within each health system were nearly matched in size. However, the cluster sizes varied across health systems (median 121, range 12-434). On average, 52.7% of the patients in each clinic were women, with a range of 36.5% to 72.4%. The most notable difference among the three fitted models (Table 7) is that the intervention effect was found to be significantly different between men and women in the GLMM (p-value: 0.01), but not significantly different in the flexible GLMM and GEE (p-values: 0.64 and 0.93, respectively). Nevertheless, the 95% confidence intervals for the PROUD intervention effects within men and women, as determined by GLMM, overlapped, consistent with the findings of the flexible GLMM and GEE. This difference may be attributed to potential heterogeneous within-cluster correlations by sex, which may have led to underestimation of standard errors. In each simulation of Scenario 3, the effect modification estimate from the GLMM was closer to the estimate from GEE, compared to the flexible GLMM's estimate; yet, all the effect modification estimates are asymptotically unbiased for their respective target estimand. Conversely, in this application we see that the effect modification estimates from the three models are slightly different. Although small, these discrepancies in the effect modification estimates among the three fitted models could be attributed to sex imbalances across clinics, informative clinic sizes or that the different approaches are estimating slightly different quantities since the GLMM methods do not weight participants equally (patients in smaller clinics have more weight than larger clinics) relative to GEE with independent working correlation does weight equally.

## 5. Discussion

This article spotlights a major analytical challenge that can arise when inferring effect modification by subgroups using clustered data. Focusing on the cluster RCT setting, we found that ignoring subgroup-specific correlations within clusters can lead to severe type I error inflation in GLMM when testing for effect modification (up to 47.2% across our simulation scenarios versus the nominal 5% level). To address this issue, we proposed a flexible GLMM that accounts for subgroup-specific correlations in addition to within-cluster correlations. In the "ideal" cluster RCT setting (e.g., having ≥ 50 clusters with non-trivial random effects) the proposed flexible GLMM controlled the type I error rate at the nominal level as did GEE (though rates were slightly elevated even with as many as 50 clusters). When there are truly no subgroup-specific correlations within clusters (i.e., the standard GLMM is the true model and the flexible GLMM is over-specified), the type I error rate under the flexible GLMM was conservatively controlled.

We also examined the impact of misspecified GLMMs under more challenging scenarios of cluster RCTs, including scenarios with few clusters or small random effects. With small random effects, the flexible GLMM conservatively controls the type I error rate and may result in a singular fit. We therefore recommend considering a two-step procedure that subsequently fits the standard GLMM model after encountering a singular fit when fitting the flexible GLMM. In scenarios with a small number of clusters, the flexible GLMM utilizing small-sample correction methods yielded type I errors closer to the nominal level compared to the standard GLMM. However, it still exhibited some inflation for binary outcomes, albeit smaller than the Type I error inflation observed with GEE. Based on our simulation study, we identified which correction method for studies with few clusters performed best for flexible and standard GLMMs when the true model is known. These correction methods demonstrated acceptable control of type I errors close to the nominal level. In practical applications, where the true correlation structure within clusters is unknown, we recommend applying the selected correction methods for both the flexible GLMM and the standard GLMM. If the results obtained using these methods differ significantly, we advise reporting results of the flexible GLMM. The various small-sample correction methods recommended for each combination of fitted and true models include quite complex and ad-hoc approaches. Such complexity underscores the necessity for future research to evaluate and propose robust methods applicable to both binary and count outcomes, irrespective of the true model. This could involve considering wild bootstrap resampling methods to obtain the empirical distribution of null effect modification{MacKinnon, 82018 #25}.

In conclusion, we recommend that studies evaluating HTE in settings with clustered data carefully consider the correlation structure when using GLMM methods. For CRTs, we recommend using GEE or a flexible GLMM that allows for separate within-cluster correlations across subgroups. Future work is needed to examine the performance of the methods in other settings of clustered data (e.g., multiple layers of clustering, and settings with imbalanced subgroups across clusters).

**Funding Statement:**

Research reported in this publication was supported by the National Institute On Drug Abuse of the National Institutes of Health under Award Number UG1DA040314 (Health Systems Node). The content is solely the responsibility of the authors and does not necessarily represent the official views of the National Institutes of Health.

# Appendix

In this section we provide example R code implementing GEE, GLMM and the flexible GLMM for a single generated dataset with a count outcome.

```
library(mvtnorm)

library(mvProbit)

library(lme4)

library(gee)

## Function of generating count data

count.gen<-function(nc,cs,beta,V){

  out<-NULL

  trtlist=c(rep(0,nc/2),rep(1,nc/2))

  trt.assign<-sample(1:nc,replace=FALSE)

  for(i in 1:nc){

    sid<-seq(1,cs)+(i-1)*cs

    trt=trtlist[trt.assign[i]]

    sex<-rbinom(n=cs,size=1,prob=c(0.5))

    ns1=sum((sex==1)+0)

    ns0=sum((sex==0)+0)

    u<-rmvnorm(n=1,mean=rep(0,2),sigma=V)

    re<-rep(NA,cs)

    re[sex==1]<-u[1]

    re[sex==0]<-u[2]

    model=exp(beta[1]+beta[2]*trt+beta[3]*sex+re)
```

```r
      y=rpois(cs,model)

      out<-rbind(out,data.frame(clinic_id=i,sid,y,trt=trt,sex))
  }
  return(out)
}

#parameters setting
#number of clusters
nc=50;
#cluster size/
cs=100;
V=symMatrix( c( 0.5,0.25,0.5 ) )
beta<-c(-1, -0.07, -0.5)

# Generating count data
sdata<-count.gen(nc,cs,beta=beta,V)
sdata$sex<-factor(sdata$sex)

# Fit GLMM
fit<-glmer(y ~ trt+sex+trt*sex+ (1|clinic_id), data=sdata, family=poisson,
       control = glmerControl(optCtrl = list(maxfun = 10000)))
```
# Fit flexible GLMM: in this code, we uses random effects for the intercept and slope of sex. This re-parametrization is equivalent to modeling subgroup-specific random intercept effects when sex are dummy variables.
```r
fit2<-glmer(y ~ 1+trt+sex+trt*sex+ (1+sex|clinic_id), data=sdata, family=poisson,
       control = glmerControl(optCtrl = list(maxfun = 10000)))
```
# Fit GEE
```r
fit3<-gee(y ~ 1+trt+sex+trt*sex,id=clinic_id,data=sdata,family="poisson")
```

*Table 1. Simulation Setting by Scenario*

| Scenario description | True model | Covariance $\Sigma$ for $(U_{i,0}, U_{i,1})$ or variance $\sigma^2$ for $U_i$ | Number of clusters, N |
|---|---|---|---|
| Scenario 1: ideal settings for flexible GLMM in terms of adequate random effect and number of clusters | Flexible GLMM | $\begin{pmatrix} 0.2 & 0.13 \\ 0.13 & 0.1 \end{pmatrix}, \begin{pmatrix} 0.5 & 0.25 \\ 0.25 & 0.5 \end{pmatrix}$ and $\begin{pmatrix} 0.25 & 0.18 \\ 0.18 & 0.5 \end{pmatrix}$ for continuous, count and binary outcomes, respectively. | N=50 and 100 with fixed cluster size 100 |
| Scenario 2: small random effect but adequate number of clusters | Flexible GLMM | Covariances are downsized by 1/10 of the variances in Scenario 1 with the same correlation degrees. These covariances reflect PROUD data application. | N=50 and 100 with fixed cluster size 100 |
| Scenario 3: adequate random effect but small number of clusters | Flexible GLMM | The same covariances of Scenario 1 | N=12 with fixed cluster 100 and N=12 with varying cluster size of (25, 50, 100, 150, 300) |
| Scenario 4: to evaluate the robustness of flexible GLMM when flexible GLMM is over fitting. | GLMM | $\sigma^2$ is 0.2, 0.5 and 0.5 for continuous, count and binary outcomes, respectively. (e.g. $\Sigma = \sigma^2 \begin{pmatrix} 1 & 0 \\ 0 & 1 \end{pmatrix}$ | N=50 with fixed cluster size 100 and N=12 with varying cluster size of (25, 50, 100, 150, 300) |

*Table 2. Selected small-sample correction methods for simulation scenarios with a small number of clusters by modeling approach and outcome type.*

| True model | Fitted model | Outcomes | | |
|---|---|---|---|---|
| | | Continuous | Count | Binary |
| Flexible GLMM | Flexible GLMM | Satterthwaite | Between-within | Subtracting the number of cluster-level parameters for fixed effects from the number of clusters |
| | GLMM | | Subtracting the number of cluster-level parameters for fixed effects from the number of clusters | Subtracting the number of cluster-level parameters for fixed effects from the number of clusters |
| GLMM | Flexible GLMM | | Parametric bootstrap | Parametric bootstrap |
| | GLMM | | Between-within | Between-within |

Table 3. Simulation results of Scenario 1 (adequate random effect and number of clusters): the inference for the null effect modification by outcomes and models

| Models | Number of clusters=50 | | | Number of clusters=100 | | |
|---|---|---|---|---|---|---|
| | Bias | ESD[a] | Type I error rate | Bias | ESD[a] | Type I error rate |
| **Continuous** | | | | | | |
| Flexible GLMM | 0.010 | 0.072 | 0.054 | 0.003 | 0.051 | 0.046 |
| GLMM | 0.003 | 0.071 | 0.218 | 0.000 | 0.051 | 0.208 |
| GEE | 0.008 | 0.075 | 0.064 | 0.000 | 0.052 | 0.050 |
| **Count** | | | | | | |
| Flexible GLMM | -0.015 | 0.237 | 0.069 | 0.001 | 0.168 | 0.059 |
| GLMM | -0.010 | 0.264 | 0.472 | 0.001 | 0.187 | 0.465 |
| GEE | -0.011 | 0.280 | 0.093 | -0.003 | 0.190 | 0.062 |
| **Binary** | | | | | | |
| Flexible GLMM | - | - | 0.062 | - | - | 0.056 |
| GLMM | - | - | 0.187 | - | - | 0.181 |
| GEE | - | - | 0.069 | - | - | 0.057 |

[a]ESD=empirical standard deviation; Flexible GLMM is the true model in the data generation; Cluster size 100 is fixed.

*Table 4. Simulation results of Scenario 2 (small random effect but adequate number of clusters): the inference for the null effect modification by outcomes and models*

| Models | Number of clusters=50 | | | Number of clusters=100 | | |
|---|---|---|---|---|---|---|
| | Bias | ESD[a] | Type I error rate | Bias | ESD[a] | Type I error rate |
| **Continuous** | | | | | | |
| Flexible GLMM | 0.002 | 0.051 | 0.037 (248[b]) | -0.001 | 0.037 | 0.036 (183[b]) |
| Two-step | 0.002 | 0.051 | 0.056 | -0.001 | 0.037 | 0.044 |
| GLMM | 0.002 | 0.051 | 0.092 (0[c]) | -0.001 | 0.038 | 0.085 (0[c]) |
| GEE | 0.002 | 0.052 | 0.053 | -0.001 | 0.038 | 0.043 |
| **Count** | | | | | | |
| Flexible GLMM | -0.002 | 0.127 | 0.055 (400[b]) | 0.001 | 0.087 | 0.039 (286[b]) |
| Two-step | -0.002 | 0.127 | 0.067 | 0.001 | 0.087 | 0.050 |
| GLMM | -0.002 | 0.127 | 0.085 (1[c]) | 0.001 | 0.087 | 0.077 (0[c]) |
| GEE | -0.002 | 0.127 | 0.060 | 0.001 | 0.087 | 0.040 |
| **Binary** | | | | | | |
| Flexible GLMM | - | - | 0.056 (278[b]) | - | - | 0.038 (218[b]) |
| Two-step | - | - | 0.053 | - | - | 0.050 |
| GLMM | - | - | 0.071 (75[c]) | - | - | 0.054 (28[c]) |
| GEE | - | - | 0.076 | - | - | 0.045 |

[a]ESD=empirical standard deviation; [b, c] =Number of singular fits out of 1,000 replicates; Flexible GLMM is the true model in the data generation; Cluster size 100 is fixed.

Table 5. Simulation results of Scenario 3 (adequate random effect but small number of clusters): the inference for the null effect modification by outcomes and models when the number of clusters is 12.

| Models | Number of clusters=12 with Fixed cluster size=100 | | | Number of clusters=12 with varying cluster sizes of (25, 50, 100, 150, 300) | | |
|---|---|---|---|---|---|---|
| | Bias | ESD[a] | Type I error rate | Bias | ESD[a] | Type I error rate |
| **Continuous** | | | | | | |
| Flexible GLMM[c] | -0.006 | 0.145 | 0.040 (159[b]) | 0.007 | 0.154 | 0.048 (181[b]) |
| Two-step[c] | -0.006 | 0.145 | 0.053 | 0.007 | 0.157 | 0.089 |
| GLMM[c] | -0.006 | 0.145 | 0.205 | 0.005 | 0.167 | 0.321 |
| GEE[d] | -0.006 | 0.151 | 0.067 | 0.005 | 0.169 | 0.072 |
| **Count** | | | | | | |
| Flexible GLMM[e] | -0.024 | 0.482 | 0.041 (17[b]) | 0.001 | 0.507 | 0.034 (30[b]) |
| Two-step[e,f] | -0.024 | 0.483 | 0.045 | -0.001 | 0.507 | 0.044 |
| GLMM[f] | -0.018 | 0.523 | 0.356 | 0.013 | 0.584 | 0.449 |
| GEE[d] | -0.019 | 0.525 | 0.099 | 0.011 | 0.586 | 0.123 |
| **Binary** | | | | | | |
| Flexible GLMM[f] | - | - | 0.060 (236[b]) | - | - | 0.068 (167[b]) |
| Two-step[f] | - | - | 0.090 | - | - | 0.144 |
| GLMM[f] | - | - | 0.138 | - | - | 0.210 |
| GEE[d] | - | -- | 0.072 | - | - | 0.087 |

[a]ESD=Empirical standard deviation; [b]=Number of singular fits out of 1,000 replicates; Flexible GLMM is the true model in the data generation; [c] = Satterthwaite; [d] = Fary and Graubard; [e] =between-within; [f] = Subtracting the number of cluster-level parameters for fixed effects from the number of clusters.

Table 6. Simulation results of Scenario 4 (flexible GLMM is over fitting): the inference for the null effect modification by outcomes and models.

| Models | Number of clusters=50 with fixed cluster size 100 | | | Number of clusters=12 with varying cluster sizes of (25, 50, 100, 150, 300) | | |
|---|---|---|---|---|---|---|
| | Bias | ESD[a] | Type I error rate | Bias | ESD[a] | Type I error rate |
| **Continuous** | | | | | | |
| Flexible GLMM | -0.001 | 0.046 | 0.045 (402[b]) | -0.002 | 0.090 | 0.032 (574[b]) |
| Two-step | -0.001 | 0.045 | 0.046 | -0.002 | 0.089 | 0.038 |
| GLMM | -0.001 | 0.046 | 0.058 | -0.002 | 0.087 | 0.046 |
| GEE | 0.000 | 0.051 | 0.055 | -0.001 | 0.097 | 0.062 |
| **Count** | | | | | | |
| Flexible GLMM | 0.001 | 0.099 | 0.036 (273[b]) | 0.002 | 0.215 | 0.031 (387[b]) |
| Two-step | 0.000 | 0.097 | 0.040 | -0.003 | 0.212 | 0.042 |
| GLMM | 0.001 | 0.096 | 0.045 | -0.002 | 0.203 | 0.051 |
| GEE | 0.002 | 0.104 | 0.063 | 0.002 | 0.214 | 0.077 |
| **Binary** | | | | | | |
| Flexible GLMM | - | - | 0.039 (173[b]) | - | - | 0.041 (349[b]) |
| Two-step | - | - | 0.039 | | | 0.043 |
| GLMM | - | - | 0.046 | - | - | 0.041 |
| GEE | - | - | 0.069 | - | - | 0.072 |

[a]ESD=empirical standard deviation; [b]=Number of singular fits out of 1,000 replicates The GLMM is the true model in the data generation; when the number of clusters is 12, the small cluster correction methods were Satterwhite for continuous outcome across GLMM and flexible GLMM, parametric bootstrap for flexible GLMM across count and binary outcomes, between-within for GLMM across count and binary outcomes and Fay-Gray for GEE across the three outcomes.

*Table 7. Comparison of three models applied to the PROUD data.*

| | Flexible GLMM | | | GLMM | | | GEE | | |
|---|---|---|---|---|---|---|---|---|---|
| Fixed effects | Beta | Standard Error | P-value | Beta | Standard Error | P-value | Beta | Standard Error | P-value |
| Intercept | -5.34 | 0.30 | <0.001[a] | -5.38 | 0.26 | <0.001[b] | -4.94 | 0.09 | <0.001[b] |
| Intervention PROUD | -0.067 | 0.43 | 0.88[a] | 0.21 | 0.37 | 0.58[b] | -0.20 | 0.25 | 0.19[b] |
| Male | -0.090 | 0.32 | 0.79[a] | 0.015 | 0.03 | 0.60[b] | -0.04 | 0.06 | 0.83[b] |
| Baseline Outcome | 0.28 | 0.003 | <0.001[a] | 0.27 | 0.003 | <0.001[b] | 0.28 | 0.02 | <0.001[b] |
| Effect modification (Inter × Male) | 0.21 | 0.45 | 0.64[a] | -0.14 | 0.04 | 0.01[b] | 0.02 | 0.14 | 0.93[b] |
| Intervention effect by subgroup | Effect (95% Confidence interval[a]) | | | Effect (95% Confidence interval[b]) | | | Effect (95% Confidence interval[b]) | | |
| Within men | 0.14 (-0.84, 1.13) | | | 0.072 (-0.75, 0.89) | | | -0.18 (-0.60, 0.24) | | |
| Within women | -0.067 (-1.02, 0.89) | | | 0.21 (-0.61, 1.03) | | | -0.20 (-0.51. 0.12) | | |

[a]=Between-within correction; [b]= Subtracting the number of cluster-level parameters for fixed effects from the number of clusters.

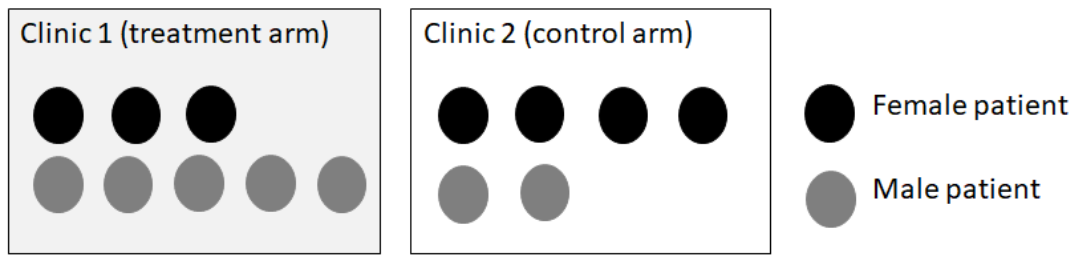

*Figure. Illustration of an example of two arms with two clusters in a typical cluster-randomized trial, in which patients are classified within sex-subgroups (here females and males are represented by black and dark grey disks, respectively) and are nested within clinic (cluster) with cluster-level treatment group assignment. In addition to clustering of patients within subgroups and clinics, the sample size varies across clinics, both overall and within subgroups.*